\documentclass[nofootinbib,prx,twocolumn,showpacs,superscriptaddress,notitlepage,superscriptaddress,amsmath]{revtex4-2}

\usepackage{listings}
\usepackage{xcolor}
\usepackage{lipsum}  
\usepackage{dsfont}
\usepackage{amsmath}
\usepackage{braket}
\usepackage{amssymb}
\usepackage{amsthm}
\usepackage{algpseudocode}
\usepackage{algorithm}
\usepackage{siunitx}
\usepackage{amsfonts}
\usepackage{comment}
\usepackage[normalem]{ulem}
\usepackage{graphicx}
\usepackage{color,framed}
\usepackage{hyperref}
\usepackage{enumerate}
\usepackage{lipsum}
\usepackage{slashed}
\usepackage{url}
\usepackage{bbm}
\usepackage{chngcntr}
\counterwithout{equation}{section}
\usepackage{tikz,pgfplots}
\usepackage{pgfplotstable}
\usepackage{siunitx}
\usepackage{comment}
\usepackage{graphicx}

\makeatletter
\newcommand{\setlabel}[1]{\edef\@currentlabel{#1}\label}
\makeatother

\usepgfplotslibrary{fillbetween}

\hypersetup{
    colorlinks=true, 
    linktoc=all,     
    linkcolor=blue,  
}

\def \beq {\begin{equation}}
\def \eeq {\end{equation}}
\def \beqa {\begin{eqnarray}}
\def \eeqa {\end{eqnarray}}
\def \bseq {\begin{subequations}}
\def \eseq {\end{subequations}}

\newcommand \tr {{\rm tr}\,}

\newcommand{\D}[1]{\text{d}#1}

\pgfplotsset{compat=1.18}
\begin{document}

\title{Analog simulation of noisy quantum circuits}
\author{Etienne Granet}\author{K\'evin H\'emery}\author{Henrik Dreyer}
\affiliation{Quantinuum, Leopoldstrasse 180, 80804 Munich, Germany}
\date{\today}
\begin{abstract}
    It is well-known that simulating quantum circuits with low but non-zero hardware noise is more difficult than without noise. It requires either to perform density matrix simulations (coming with a space overhead) or to sample over  ``quantum trajectories" where Kraus operators are inserted randomly (coming with a runtime overhead). We propose a  simulation technique based on a representation of hardware noise in terms of trajectories generated by operators that remain close to identity at low noise. This representation significantly reduces the variance over the quantum trajectories, speeding up noisy simulations by factors around $10$ to $100$. As a by-product, we provide a formula to factorize multiple-Pauli channels into a concatenation of single Pauli channels.
    
\end{abstract}
\maketitle

\section{Introduction}
While quantum computers's capacities are steadily improving, both in qubit number or in gate fidelity \cite{moses2023race,decross2024computational,kim2023evidence}, application-oriented quantum engineers will likely have to deal with hardware noise for years to come. Besides the exact simulation of hardware on a small number of qubits, their \emph{noisy} simulation is thus of crucial importance to optimize the noise mitigation part of an algorithm implementation. However, it is well-known that simulating quantum circuits with low but non-zero noise is more difficult than without noise \cite{xu2023herculean}. Indeed, it requires either to perform density matrix simulations (coming with a space overhead) or to sample over  ``quantum trajectories" where Kraus operators are inserted randomly (coming with a runtime overhead). For example, while $\sim 25$ noiseless qubits can be routinely simulated on a laptop, $\sim 25$ noisy qubits take considerably longer with some widely used tools \cite{qiskit2024,cirq_developers_2024_11398048,cudaq,isakov2021simulations}. We note that larger noise strength can become easier to simulate than noiseless with other techniques \cite{tindall2024efficient,liao2023simulation,beguvsic2024fast,ayral2023density,fontana2023classical,cheng2021simulating}, but these very noisy regimes are less interesting to the quantum engineer.

The objective of this work is to present a classical simulation method of noisy quantum circuits that yield $1$ to $2$ orders of magnitude of runtime speedup. The basic principle is to express noisy expectation values as sums over quantum trajectories, but these trajectories are generated in a different, more optimal way than with Pauli Kraus operators insertion. Instead of inserting e.g. a $X$ error with a probability $\epsilon$, one systematically inserts a small random rotation $e^{i\theta X}$ with an angle $\theta$ whose variance scales as $\epsilon$. The variance of expectation values over different trajectories tends to be much smaller with this approach, which reduces the number of trajectories to simulate. Intuitively, trajectories display smaller statistical fluctuations if one introduces systematic but small perturbations after every gate rather than rare but large perturbations. One notes that this simulation technique does not increase the class of classically simulable circuits. But it provides instead a significant speedup of noisy simulations routinely performed by quantum practitioners.

As a by-product of this work, we also obtain a formula for factorizing an arbitrary Pauli channel with multiple Pauli strings into the concatenation of multiple Pauli channels with only one Pauli string.

Different ways of ``unravelling" noise channels into quantum trajectories have been considered in the past \cite{dalibard1992wave,molmer1993monte,carmichael2009open,plenio1998quantum} and more recently \cite{chen2023optimized,ferris2012variational,vovk2022entanglement,kolodrubetz2023optimality,di2023noisy} mainly focusing on reducing the entanglement in each trajectory. It seems however that no simple protocol to reduce the variance over trajectories has been proposed, as evidenced by the fact that, as far as we know, all simulation packages publicly available such as Qiskit or Cirq implement the usual Kraus operator insertions \cite{qiskit2024,cirq_developers_2024_11398048,cudaq,isakov2021simulations}.

\begin{figure}
    \centering
    \includegraphics[width=\linewidth]{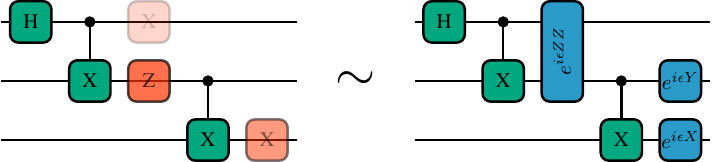}
    \caption{Gates (green boxes) after which Kraus operators are randomly inserted (red boxes) are equivalent in average to gates after which rotations with a small random angle $\epsilon$ are systematically inserted (blue boxes).}
    \label{conceptualfigure}
\end{figure}

\section{Simulating noisy hardware}
\subsection{Generalities}
We consider $N$ qubits on which we would like to apply a unitary operator $U$. If the qubits are initialized in a pure state $|\psi\rangle$, the outcome of the operation on a perfect, noiseless quantum hardware is $U|\psi\rangle$. This requires  to store $2^N$ numbers to simulate classically. 

On noisy quantum hardware, the $N$ qubits have to be described by a density matrix $\rho$. The unitary $U$ acts on this density matrix as $\mathcal{U}(\rho)=U \rho U^\dagger$. Let us denote by $\mathcal{U}_{\rm noisy}$ the same operation performed on a noisy hardware. It is well-known that it can be generically written as \cite{nielsen2010quantum}
\begin{equation}
    \mathcal{U}_{\rm noisy}(\rho)=\mathcal{N}(\mathcal{U}(\rho))\,,
\end{equation}
with
\begin{equation}\label{genericN}
    \mathcal{N}(\rho)=(1-\epsilon)\rho+\epsilon \sum_{q=1}^Q K_q \rho K_q^\dagger\,,
\end{equation}
with $0 \leq \epsilon \leq 1$, $Q$ is an integer and $K_q$ Kraus operators satisfying
\begin{equation}
    \sum_{q=1}^Q K_q^\dagger K_q ={\rm Id}\,.
\end{equation}
We note that the writing \eqref{genericN} also applies to non-Pauli channels like amplitude damping by allowing $\epsilon=1$. Using this expression to simulate the noisy circuit by evolving the density matrix $\rho$ comes with a significant space overhead compared to the noiseless simulation as this requires storing and updating $4^N$ numbers.

\subsection{Trajectory-based noise simulation}
\subsubsection{Definition}
A \emph{trajectory-based} noisy simulation consists in averaging randomly generated \emph{pure states} according to a well-chosen random process, in such a way that their corresponding density matrices statistically average to $\rho$ \cite{dalibard1992wave,molmer1993monte,carmichael2009open,plenio1998quantum}. Equivalently, this means generating random operators $W$ in such a way that for any state $|\psi\rangle$ we have
\begin{equation}
   \mathcal{N}(|\psi\rangle\langle\psi|)=\mathbb{E}[W|\psi\rangle \langle \psi|W^\dagger]\,, 
\end{equation}
where $\mathbb{E}$ denotes the statistical average over the operators $W$. Expectation values of an operator $O$ within the noisy state $\rho=\mathcal{N}(|\psi\rangle\langle\psi|)$ are thus obtained as
\begin{equation}
    \tr[\rho O]=\mathbb{E}[\langle \psi|W^\dagger O W|\psi\rangle]\,.
    \label{eq:sampling}
\end{equation}
We note that the $W$ operators do not need to be unitary. In case they are not unitary, the state used in the simulation can change norm.

In terms of space, this approach is as costly as performing state-vector noiseless simulations. However, the averaging over multiple trajectories comes with a sampling overhead which increases the total runtime of the simulation. In order to reach precision $\eta$ on $\tr[\rho O]$, one needs to compute
\begin{equation}
    M=\frac{V}{\eta^2}
\end{equation}
different trajectories, with the variance $V$ over the trajectories
\begin{equation}
    V=\mathbb{E}[\langle \psi|W^\dagger O W|\psi\rangle^2]-\mathbb{E}[\langle \psi|W^\dagger O W|\psi\rangle]^2\,.
\end{equation}
The runtime of the noisy simulation is thus proportional to the variance $V$.

\subsubsection{Example: digital sampling \label{sec:digital}}
The most widely used trajectory sampling protocol in noisy circuit simulation is the following \cite{qiskit2024,cirq_developers_2024_11398048}, which we will refer to as ``digital" sampling. One sets $K_q=\sqrt{p_q}\bar{K}_q$ with the norm-2 $||\bar{K}_q||=1$, and $p_q$ some number. The random operator $W$ is then defined as
\begin{equation}
    W=\begin{cases}
        I\,,\qquad \text{with probability}\quad 1-\epsilon\\
        \bar{K}_q\,,\qquad \text{with probability}\quad \epsilon \frac{p_q}{\sum_{q'=1}^Qp_{q'}}\,.
    \end{cases}
\end{equation}
This trajectory sampling protocol is popular as it is simple to implement and provides physical intuition to the noise channel \eqref{genericN}: the Kraus operators are seen as ``errors" occurring with a certain probability after application of each gate. It is also the trajectory sampling used in the theory of quantum error correction: if the noise channel \eqref{genericN} is a Pauli channel, then all errors can be seen as applying $X$ and/or $Z$ errors on the qubits. 

However, in terms of classical simulation of noisy circuits, this trajectory sampling has a large variance and is highly suboptimal. This can be understood intuitively. For a low-noise circuit, most of the time, the operator $W$ inserted after every gate is just identity. However, when an error occurs, which happens rarely, $W$ is a completely different operator that modifies the state significantly. The time evolution of operators along the trajectories will thus display ``jumps" at insertion of errors and two distinct trajectories will be very dissimilar. A better sampling strategy would be to systematically insert a small perturbation after every gate, instead of rarely inserting a large perturbation. Intuitively, this suggests to consider random operators $W$ that always differ from identity, but remain close to it $||W-I||=\mathcal{O}(\epsilon)$.

\section{Analog sampling}
\subsection{Digital vs analog sampling}
We consider a parameter-dependent noise channel $\mathcal{N}_\epsilon$ such that for all $\rho$ 
\begin{equation}
    ||\mathcal{N}_\epsilon(\rho)-\rho||=\mathcal{O}(\epsilon)\,,
\end{equation}
with $\epsilon$ controlling the amplitude of the noise, and $W_\epsilon$ a random operator such that for all $|\psi\rangle$
\begin{equation}
    \mathbb{E}[W_\epsilon |\psi\rangle\langle\psi|W_\epsilon^\dagger]=\mathcal{N}_{\epsilon}(|\psi\rangle\langle\psi|)\,.
\end{equation}
For $\eta>0$, we define then $\tilde{W}_\epsilon^\eta$ the random operator equal to $W_\epsilon$ if $||W_\epsilon-I||<\eta$, and equal to $I$ otherwise. We say that $W_\epsilon$ is an \emph{analog representation} of the noise channel $\mathcal{N}_\epsilon$ if for all $\eta>0$ and density matrix $\rho$, we have when $\epsilon\to 0$
\begin{equation}\label{conditionanalog}
     \mathbb{E}[\tilde{W}^\eta_\epsilon \rho \tilde{W}^{\eta\dagger}_\epsilon ]=\mathcal{N}_\epsilon(\rho)+o(||\rho-\mathcal{N}_\epsilon(\rho)||)\,.
\end{equation}
Loosely speaking, an analog representation of a noise channel is a quantum trajectory where all extra operators inserted are close to identity.

The digital representation presented in Section \ref{sec:digital} does \emph{not} satisfy this condition in general. For example, for a Pauli channel with Kraus operator $K_1=X$, since $||X-I||=2$, for any $\eta<2$ we have always $\tilde{W}^\eta_\epsilon=I$, and hence \eqref{conditionanalog} cannot hold. When \eqref{conditionanalog} does not hold, it means that the trajectory sampling has to insert operators at distance $\mathcal{O}(1)$ from $I$, and so will make the trajectories vary significantly from one sample to another.

In the following subsections, we will present some simple ways of implementing analog noise representation of different quantum channels.

\subsection{Single Pauli string channel \label{analogsingle}}
Let us consider first a Kraus operator $K$ that is a product of Pauli matrices, and define a noise channel $\mathcal{N}_{q,K}$ containing only one Kraus operator as
\begin{equation}
    \mathcal{N}_{q,K}(\rho)=(1-q)\rho +q K \rho K\,.
\end{equation}
Let us apply on $\rho$ a rotation $e^{i\theta K}$ with probability density $f(\theta)$. Using $K^2=I$, the density matrix that we obtain is
\begin{equation}\label{analogsingle0}
    \int_{-\infty}^\infty e^{i\theta K}\rho e^{-i\theta K} f(\theta)\D{\theta}=a \rho+b K\rho K+ic(K\rho-\rho K)\,,
\end{equation}
with
\begin{equation}
    \begin{aligned}
        a=&\int_{-\infty}^\infty \cos^2 (\theta) f(\theta)\D{\theta}\,,\qquad b=\int_{-\infty}^\infty \sin^2 (\theta) f(\theta)\D{\theta}\\
        c=&\int_{-\infty}^\infty \cos (\theta)\sin (\theta) f(\theta)\D{\theta}\,.
    \end{aligned}
\end{equation}
Since $\int f(\theta)\D{\theta}=1$, we have $0\leq a\leq 1$, as well as $b=1-a$. Hence, if the probability $f(\theta)$ is chosen such that $c=0$, then we have
\begin{equation}\label{angle}
    \int_{-\infty}^\infty e^{i\theta K}\rho e^{-i\theta K} f(\theta)\D{\theta}=\mathcal{N}_{q,K}(\rho)\,,
\end{equation}
with $q=b$. A simple way to implement $c=0$ is just to choose $f(\theta)$ symmetric $f(\theta)=f(-\theta)$. Let us present two simple examples. One can take a Gaussian
\begin{equation}
    f(\theta)=\frac{e^{-\tfrac{\theta^2}{2\sigma^2}}}{\sqrt{2\pi\sigma^2}}\,,
\end{equation}
with the variance
\begin{equation}\label{sigmavariance}
    \sigma^2=-\frac{1}{2}\log(1-2q)\,.
\end{equation}
Another possibility is a discrete, Bernoulli random variable
\begin{equation}
    f(\theta)=\frac{1}{2}(\delta(\theta-\arcsin\sqrt{q})+\delta(\theta+\arcsin\sqrt{q}))\,.
\end{equation}
We will come back on the choice of probability distribution below in Section \ref{sec:optimalangle}. These two choices are analog representation of the noise, as the operators $e^{i\theta K}$ become arbitrarily close to identity when $q\to 0$.

\subsection{Multiple Pauli string channel}
We now would like to implement a generic noise channel like \eqref{genericN} with $Q\geq 2$, but in which the Kraus operators $K_q$ are constrained to be proportional to Pauli strings. Namely we define
\begin{equation}\label{multiplekraus}
    \mathcal{N}(\rho)=\sum_{S\in \mathcal{S}_M} p_S S \rho S\,,
\end{equation}
with $\mathcal{S}_M$ the set of all $4^M$ Pauli strings on some fixed $M$ qubits, and $0\leq p_S\leq 1$ some probabilities of error satisfying $\sum_{S\in \mathcal{S}_M} p_S=1$. Generalizing the analog noise representation to multiple non-trivial Kraus operators is not trivial. We present two approaches to do so.

\subsubsection{Factorization approach}
One first approach is to ``factorize" the noise channel \eqref{multiplekraus} into single-Kraus-operator channels, as
\begin{equation}\label{multiplefactorized}
    \mathcal{N}(\rho)=\left[\prod_{S\in\mathcal{S}_M}\mathcal{N}_{q_S,S}\right](\rho)\,.
\end{equation}
Here, the product of the maps $\mathcal{N}_{q_S,S}$ means their composition. For example, for $M=2$ this product includes $15$ single-Kraus-operator channels
\begin{equation}
    \mathcal{N}(\rho)=\mathcal{N}_{q_{XI},XI}(\mathcal{N}_{q_{YI},YI}(\mathcal{N}_{q_{ZI},ZI}(\mathcal{N}_{q_{XX},XX}(...))))\,.
\end{equation}
We note that since two Pauli strings always either commute or anticommute, all the channels $\mathcal{N}_{q_S,S}$ commute and the ordering in the expression above has no influence. This factorization is non-trivial, and it is a priori a complex problem to express the $q_S$ in this expression in terms of the $p_S$ in the expression \eqref{multiplekraus}, as the relation between the two sets of probabilities is non-linear. As proved in Appendix \ref{appendixa}, we obtain the following formula
 \begin{equation}\label{qQ}
    q_S=\frac{1}{2}-\frac{1}{2}\left(\frac{\prod_{S'\in a(S)} \left[1-2\sum_{S''\in a(S')}p_{S''}\right]}{\prod_{S'\in c(S)} \left[1-2\sum_{S''\in a(S')}p_{S''}\right]}\right)^{2/4^M}\,,
\end{equation}
where $a(S)$/$c(S)$ denotes the subset of Pauli strings in $\mathcal{S}_M$ that anticommute/commute with $S$.

We note that $q_S$ given by this formula is not necessarily positive. If one of the $q_S$ is not positive, then it means that the noise channel \eqref{multiplekraus} cannot be factorized into physical error channels as in \eqref{multiplefactorized} with positive probabilities. The factorization remains possible if one allows for arbitrary $q_S$. However, we expect on physical grounds that noise channels describing actual physical devices will always be factorizable into physical error channels. Indeed, hardware noise modelled by for example a Pauli channel on two qubits, containing $15$ different Pauli string errors, results from multiple physical sources such as stray magnetic fields, spontaneous emission events, cross-talks, that are each associated with a different stage of the realization of a quantum gate, and as such are more naturally represented by a composition of different simpler channels. If all the $q_S$ are positive, then \eqref{multiplefactorized} can be used to implement the channel \eqref{multiplekraus}, by using the analog sampling of Section \ref{analogsingle} for every $\mathcal{N}_{q_S,S}$.

The factorization or divisibility of quantum channels in different ways has been studied extensively in the literature \cite{wolf2008dividing}. We are not aware of previous quoting of formula \eqref{qQ} for the factorization \eqref{multiplefactorized}, which is a new result in itself.\\

Let us illustrate formula \eqref{qQ} with the simple example of a depolarizing channel on $M$ qubits with amplitude $\epsilon$. This corresponds to taking $p_S=\frac{\epsilon}{4^M}$ for $S\in\mathcal{S}_M$ different from identity, and $p_I=1-\epsilon \frac{4^M-1}{4^M}$. As explained in Appendix \ref{appendixa}, given a Pauli string $S\neq I$, there are always $4^M/2$ Pauli strings $S'\in a(S)$ that anticommute with it, and all these strings are different from $I$. For $S'=I\in c(S)$, we have $a(S')=\emptyset$. We thus find for $S\neq I$
\begin{equation}
    q_S=\frac{1}{2}-\frac{1}{2}(1-\epsilon)^{2/4^M}\,.
\end{equation}
Implementing the analog trajectories with a Gaussian distribution, this leads to the variance
\begin{equation}
    \sigma^2=-\frac{1}{4^M}\log(1-\epsilon)\,
\end{equation}
for each of the angles of the $4^M-1$ rotations to apply. Examples of application of this simulation approach with sample codes are provided in Appendix \ref{sec:samplecode}.

\subsubsection{Randomly sampled rotation approach \label{semianalog}}
Another possible approach is to first pick a string $S\in\mathcal{S}_M$ different from identity $S\neq I$ with probability $p'_S$ given by
\begin{equation}
    p'_S=\frac{p_S}{\sum_{\substack{T\in \mathcal{P}_M\\ T\neq I}} p_T}\,.
\end{equation}
Then, we apply the analog noise representation of Section \ref{analogsingle} for a single Kraus operator $K=S$ with $q=\sum_{\substack{T\in \mathcal{P}_M\\ T\neq I}} p_T$. The advantage of this approach is that it would apply even in cases where the factorization \eqref{qQ} does not lead to physical positive probabilities $q_S$. However, we expect it to be slightly suboptimal in general compared to the factorization approach.

\subsection{Non-Pauli string channel \label{non-pauli}}
We now generalize the analog representation of Section \ref{analogsingle} to non-Pauli noise.
\subsubsection{Coherent noise}
We first consider coherent noise, in the form of a single Kraus operator implementing an over-rotation $K_1=e^{i\alpha X}$ with $\alpha=\mathcal{O}(1)$, occurring with probability $q$. We consider only one such biased rotation, as including another symmetric rotation $K_2=e^{-i\alpha X}$ with same probability would make the channel equivalent to a Pauli $X$ channel \cite{nielsen2010quantum}. As detailed in Appendix \ref{app:coh}, this coherent noise channel can be implemented by applying a rotation $e^{i\theta X}$ with probability density $f(\theta)$ given by
\begin{equation}
    f(\theta)=\frac{e^{-\frac{(\theta-\mu)^2}{2\sigma^2}}}{\sqrt{2\pi \sigma^2}}\,,
\end{equation}
with
\begin{equation}
\begin{aligned}
    &\mu=\frac{1}{2}\arctan \frac{q\sin(2\alpha)}{1-2q\sin^2\alpha}\\
    &\sigma^2=-\frac{1}{4}\log (1-4q(1-q)\sin^2\alpha)\,.
\end{aligned}
\end{equation}
At small error probability, both mean and variance of the angles go to $0$. It is thus an analog representation of this coherent noise channel.

\subsubsection{Amplitude damping}
Amplitude damping is a non-Pauli noise channel that can be represented by the two Kraus operators
\begin{equation}
    K_1=\left(\begin{matrix}
        1&0\\
        0&\sqrt{1-\gamma}
    \end{matrix}\right)\,,\quad K_2=\left(\begin{matrix}
        0&\sqrt{\gamma}\\
        0&0
    \end{matrix}\right)\,,
\end{equation}
where $0\leq \gamma\leq 1$ is a parameter \cite{nielsen2010quantum}. In that case, we would take $\epsilon=1$ in the generic writing \eqref{genericN} and see the parameter $\gamma$ as tuning the amplitude of the noise, from $\gamma=0$ (no damping) to $\gamma=1$ (complete damping). A digital trajectory of this noise channel would be to apply $K_j$ on the state with probability $\langle \psi|K_j^\dagger K_j |\psi\rangle$, where $|\psi\rangle$ is the current state, and then to normalize the state obtained. This would have high variance as it would lead to jumps when an operator $K_2$ is applied. Instead, we can define the following analog trajectory. Using the fact that $K_2^2=0$ and $K_1K_2=K_2$, we have for $\theta$ an angle with probability $f(\theta)$
\begin{equation}
\begin{aligned}
     K_1e^{i\theta K_2}\rho e^{-i\theta K_2^\dagger}K_1&=K_1\rho K_1+\theta^2 K_2 \rho K_2^\dagger\\
     &+i\theta(K_2\rho K_1-K_1\rho K_2)\,.
\end{aligned}
\end{equation}
It follows that by choosing $f(\theta)$ such that $\int \theta f(\theta)\D{\theta}=0$ and $\int \theta^2 f(\theta)\D{\theta}=1$, we obtain indeed an amplitude damping channel after statistically averaging over the trajectories. The operators that we apply, $K_1$ and $e^{i\theta K_2}$, are then close to identity when $\gamma\to 0$. This is thus an analog representation of amplitude damping. We note however that the operator $e^{i\theta K_2}$ is not unitary, and so the norm of the state can change along the trajectory.

\subsection{Optimal angle distribution \label{sec:optimalangle}}
Let us investigate the effect of the angle distribution on the variance over trajectories in a simple toy model. We consider a single qubit prepared in $|0\rangle$, on which we apply repeatedly the $Z$ operator $n$ times, with a noise model where a $X$ error is inserted with probability $q$ after every $Z$ gate. After $n$ applications of $Z$, the noiseless expectation value of $Z$ is $1$. The noisy expectation value $\langle Z\rangle_{\rm noisy}$ is
\begin{equation}
\begin{aligned}
    \langle Z\rangle_{\rm noisy}&=\sum_{k=0}^n \binom{n}{k} 
q^k (1-q)^{n-k} (-1)^k\\
&=(1-2q)^n\,.
\end{aligned}
\end{equation}
Let us now consider an analog representation of the noise. We perform a $e^{i\theta X}$ rotation after every $Z$ gate with a symmetric probability distribution $f(\theta)$, such that $\int \sin^2\theta f(\theta)\D{\theta}=q$, i.e. $\mathbb{E}[\sin^2\theta]=q$. Given a realization $\theta_0,...,\theta_{n-1}$ of these angles, the state prepared is
\begin{equation}
    |\psi\rangle=\cos \alpha |0\rangle+i\sin\alpha |1\rangle\,,\qquad \alpha=\sum_{k=0}^{n-1}(-1)^{n-1-k} \theta_k\,,
\end{equation}
and so $\langle \psi| Z |\psi\rangle=\cos(2\alpha)$. Using the independence of the $\theta$'s and the fact that they have a symmetric distribution around $0$, it follows that we have
\begin{equation}
\begin{aligned}
    \mathbb{E}[\langle \psi| Z |\psi\rangle]&=\mathbb{E}[e^{2i\theta}]^n=\mathbb{E}[\cos(2\theta)]^n=(1-2q)^n\,,
\end{aligned}
\end{equation}
as expected, i.e. the average over trajectories reproduces the noise channel. As for the variance over trajectories, this is
\begin{equation}
\begin{aligned}
    V&=\mathbb{E}[\langle \psi| Z |\psi\rangle^2]-\mathbb{E}[\langle \psi| Z |\psi\rangle]^2\\
    &=\frac{1}{2}+\frac{1}{2}\mathbb{E}[\cos(4\theta)]^n-(1-2q)^{2n}\\
    &=\frac{1}{2}+\frac{1}{2}(1-8q+8\mathbb{E}[\sin^4\theta])^n-(1-2q)^{2n}\,.
\end{aligned}
\end{equation}
It follows that for $q<1/8$, the distribution that minimizes the variance, and so that minimizes the runtime of the noisy simulation, is such that $\mathbb{E}[\sin^4\theta]$ is minimal, with the condition $\mathbb{E}[\sin^2\theta]=q$. This is equivalent to minimizing the fourth moment of $f(\arcsin\theta)/\sqrt{1-\theta^2}$ at fixed variance. This is achieved by the discrete distribution
\begin{equation}
    f(\theta)=\frac{1}{2}(\delta(\theta-\arcsin\sqrt{q})+\delta(\theta+\arcsin\sqrt{q}))\,.
\end{equation}

\subsection{Number of shots per trajectory}
Let us comment further on the equivalence between digital and analog sampling. As said in Section \ref{analogsingle}, this is manifested by the equality between the two noise channels
\begin{equation}
    (1-q)\rho+q K\rho K=\frac{1}{\sqrt{2\pi\sigma^2}}\int_{-\infty}^\infty e^{-\theta^2/(2\sigma^2)}e^{i\theta K}\rho e^{-i\theta K}\D{\theta}\,,
\end{equation}
with $\sigma$ given in \eqref{sigmavariance}. The physical process leading to the left-hand side is interpreted as a bit flip occurring with  probability $q$, whereas the physical process leading to the right-hand side is interpreted as e.g. some random fluctuating magnetic field. The equality between the two channels means that on the quantum computer, the two processes are fundamentally indistinguishable \cite{nielsen2010quantum}. In particular, both yield the same variance when averaging multiple shots. Classically, this means that if one computes only one shot per simulated trajectory, both digital and analog sampling yield exactly the same variance over shots. The speedup of analog sampling over digital sampling for classical simulation only stems from the ability of performing multiple shots per simulated trajectory (or directly compute expectation values, i.e. performing an infinite number of shots), which is not possible on a quantum computer.

\begin{figure*}
    \centering
    \includegraphics[width=0.32\linewidth]{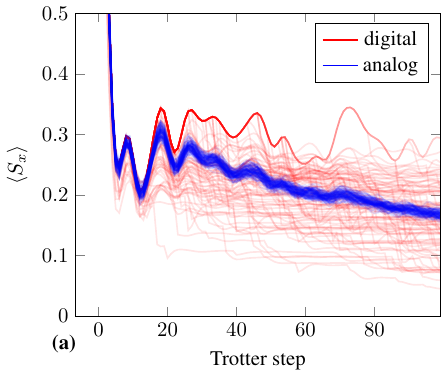}
    \includegraphics[width=0.32\linewidth]{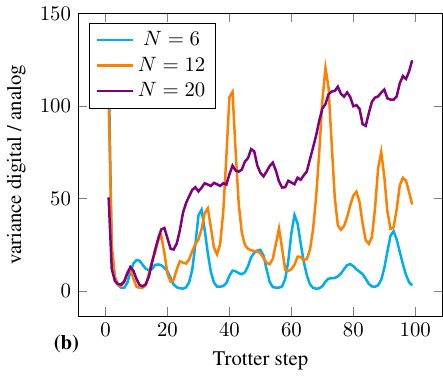}
    \includegraphics[width=0.32\linewidth]{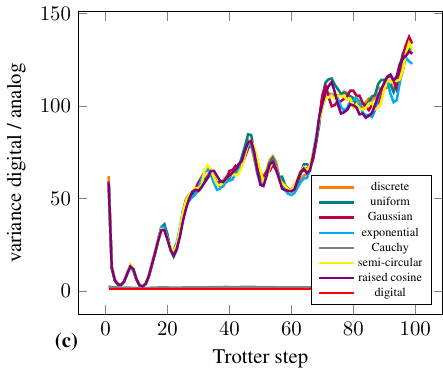}
    \caption{\textbf{(a)}: Digital trajectories (red) and analog trajectories with Gaussian distribution (blue), for a 2D Ising model on a $4\times 5$ square lattice, with $h=1$, Trotter step ${\rm d}t=0.1$, depolarizing noise $0.001$ after every two-qubit gate. \textbf{(b)}: ratio of variance of digital trajectories over variance of analog trajectories, in same setting as panel (a), for different system sizes, averaging over $700$ trajectories. This corresponds to the runtime ratio of the two noise simulations. \textbf{(c)}: same as panel (b) for $N=20$, comparing different angle probability distributions, averaging over $2000$ trajectories.}
    \label{fig:analog}
\end{figure*}

\section{Numerical implementation}
We now present numerical implementation of the analog noise representation and comparison with the usual digital approach. 
\subsection{Expectation values\label{sec:exp}}

We first consider the efficiency of the analog approach to compute expectation values of local observables in noisy circuits. As a simple and widespread benchmark, we consider Hamiltonian simulation of a 2D square lattice Ising model with Hamiltonian
\begin{equation}
    H=\sum_{\langle j,k\rangle}Z_jZ_k+h\sum_{j=1}^N X_j\,,
\end{equation}
where $h=1$ is a magnetic field, and where $\langle j,k\rangle$ means that sites $j,k$ are neighbours on the lattice. We impose periodic boundary conditions. We prepare initially the system in the state $|+...+\rangle$, implement $e^{iHt}$ using a first-order Trotterization with a Trotter step ${\rm d}t=0.1$, and measure the observable $S_x=\frac{1}{N}\sum_{j=1}^N X_j$. The noise is modeled by a two-qubit depolarizing channel with amplitude $0.001$ after every $ZZ$ rotation.

We report in Fig \ref{fig:analog} the results of the simulation. Firstly, in panel (a) we compare individual quantum trajectories using the digital simulation (red) and analog simulation (blue). The exact noisy evolution of the magnetization would be obtained by averaging all the curves. There is a stark difference between the two approaches: Digital trajectories fluctuate over a large range of values and display jumps when a Kraus operator is inserted. Single digital trajectories offer only imprecise information on the exact noisy curve. In contrast, analog trajectories are strongly localized in a ``tube" around their average value. A \emph{single} analog trajectory already correctly captures the main behaviour of the noisy expectation value.

In panel (b), we plot then the ratio of the variance of digital trajectories over the variance of analog trajectories. This ratio exactly corresponds to the ratio of the number of trajectories that one has to average over to reach a certain precision, i.e. the ratio of runtime of the two approaches. We observe that on the circuit depth and system sizes studied, the analog simulation delivers a $10$ to $100$-fold speedup over the digital approach. Moreover, this speedup is seen to increase with the number of qubits.

In panel (c), we investigate the efficiency of different probability distribution for the angles entering the analog simulation. The precise probability distributions are spelled out in Appendix \ref{sec:distributionangle}. Our toy model of Section \ref{sec:optimalangle} suggested that the discrete distribution would have the best speedup. We see here that all the distributions considered, except for the Cauchy distribution, give essentially the same variance.

Finally, in Appendix \ref{sec:samplecode}, we provide actual runtime comparisons of the digital and analog sampling. We observe again speedups between $\sim 10$ and $\sim 100$ depending on the cases considered.

\subsection{Sampling \label{sec:sampling}}
\begin{figure}
    \centering
    \includegraphics[width=\linewidth]{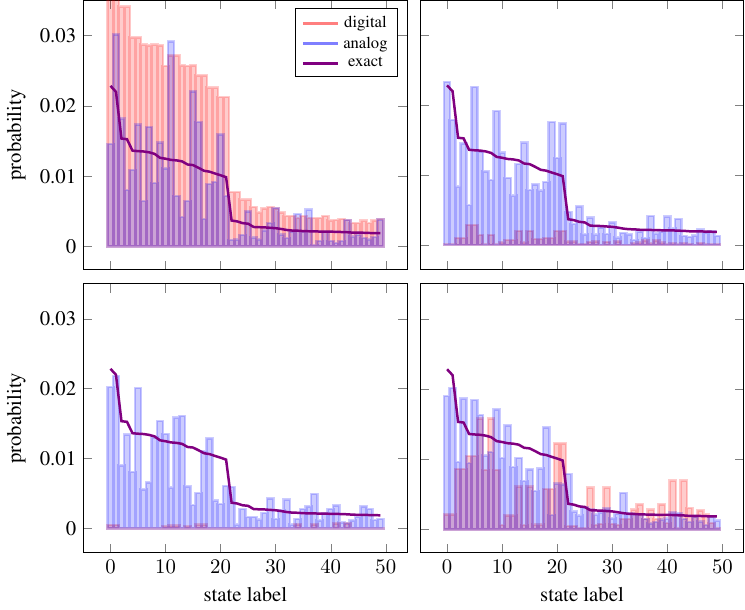}
    \caption{Comparison of probability distribution of output bitstrings for single individual trajectories generated with the digital method (red bars), analog method (blue bars), and for the exact distribution (purple line), in the setup described in Section \ref{sec:sampling}. The four panels correspond to four different randomly generated trajectories for digital and analog. The $50$ states considered are the ones with largest probabilities in the exact noisy density matrix.}
    \label{fig:histogram}
\end{figure}
Another common task of quantum computers is to perform \emph{sampling} from a wave-function in a given basis. This is particularly relevant for application of quantum computing to classical optimization problems, where the output of the algorithm can be e.g. a list of vertices of a graph. As a benchmark, we consider the Max-Cut problem on a given random $3$-regular graph on $N=20$ sites, which corresponds to finding the ground state of the classical Hamiltonian $\sum_{\langle j,k\rangle}Z_jZ_k$. We implement a ``Floquet" adiabatic evolution consisting in applying the operator
\begin{equation}
    U(s)=e^{i{\rm d}t(1-s)\sum_{j=1}^NX_j}e^{i{\rm d}ts\sum_{\langle j,k\rangle}Z_jZ_k}\,,
\end{equation}
for $s=1/T,2/T,...,1$ with $T$ an integer and ${\rm d}t>0$ a parameter \cite{granet2024benchmarking,montanez2024towards}. We initialize the system in $|+...+\rangle$, fix $T=40$ and ${\rm d}t=0.25$, and include a depolarizing channel with amplitude $0.001$ after every $ZZ$ rotation. We compute in the end the probability of sampling a given string of $N$ bits in the $Z$ basis. We report in Fig \ref{fig:histogram} the results of the numerics. We compute the ``exact" output distribution of the noisy circuits by averaging $500$ random analog trajectories. We consider the $50$ more likely bitstrings that we order by decreasing probability, and compare with the probability distribution of \emph{individual} trajectories, both digital and analog. The four different panels correspond to four different random trajectories. We see that the probability amplitudes for single analog trajectories essentially display the exact distribution, dressed by small amounts of noise. In contrast, digital trajectories display very dissimilar probability distributions. We see that one single analog trajectory can be used to approximately sample from the exact distribution, up to some noise, whereas with the digital approach one has inevitably to average over several trajectories. When computing the average absolute value of the Kullback-Leibler divergence between the exact and single trajectory distribution computed on the $50$ most likely states, we find an average value of $0.053$ for analog trajectories, compared to $0.90$ for digital trajectories.

\begin{figure*}
\includegraphics[width =0.32 \linewidth]{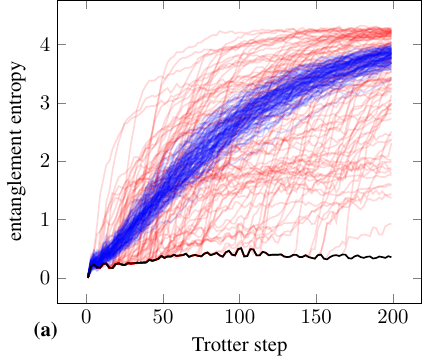}
\includegraphics[width=0.32 \linewidth]{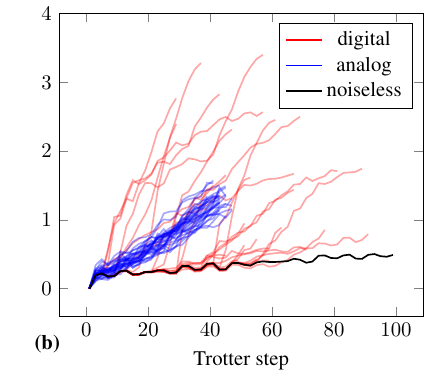}
\caption{Entanglement entropy as a function of the number of Trotter steps for the tilted field Ising model for different system sizes. We plot the exact Trotter time evolution as well as the simulated noisy trajectories for a depolarizing noise with parameter p=0.003 with both the Gaussian analog and the digital methods. \textbf{(a)} Exact results for $N=10$ \textbf{(b)} Approximate simulations with matrix product states with $\chi=32$. All the data shown on the plot have an estimated fidelity equal or above 0.95. 
}
 \label{fig:ent}
\end{figure*}

\subsection{Matrix-product-state simulation}
While so far we considered only exact state-vector simulations, these are limited to small system of the order of $25 \sim 35$ qubits. To extend the system size, tensor networks are one of the methods of choice \cite{Vidal2003,Gray2021,zhou2020limits}. These techniques however are limited by the amount of \emph{entanglement entropy} generated by the quantum circuit one wishes to simulate, on which the simulation cost depends exponentially. The growth of the entanglement entropy depends on the specific circuit structure, but is typically linear in time for generic, chaotic quantum systems and can be even larger for unstructured circuits.
Similarly to statevector simulations, the use of analog sampling has the potential to drastically reduce the simulation cost, provided the entanglement entropy of the sampled pure states $W|\psi \rangle$ remains low enough.

In Fig \ref{fig:ent}, we compare the entanglement entropy between the first and second halfs of the system when simulating a noisy Hamiltonian simulation circuit with the digital or analog sampling. Specifically, we consider the Trotterized tilted field Ising model \cite{kim_2013}:
\begin{equation}
    H= \sum_{j=1}^{N-1} Z_j Z_{j+1} + h_x\sum_{j=1}^N X_j + h_z \sum_{j=1}^N Z_j,
\end{equation}
with $h_x = 0.9045$ and $h_z=0.8090$ with a large Trotter step of ${\rm dt}=0.3$. For a generic initial product state the time evolution of this model would display a linear growth of entanglement. We observed however that if we start our simulation from the all zero state, entanglement entropy remains low, and the noiseless time evolution can be simulated up to long times. We compare then the simulation with the addition of a depolarizing noise with parameter $\lambda = 0.003$ with both digital and analog trajectory sampling methods.

In the left panel, we consider first a small system size $N=10$, for which the simulation can be carried out exactly, and thus the evolution of the entanglement entropy can be tracked up to long times. We observe that for the digital simulation, the introduction of a single error causes a sudden increase in the entanglement entropy, with some trajectories quickly saturating to the Page value. Overall, the distribution of the entanglement is very broad, indicating that some trajectories will be much more difficult to simulate than others. 
In contrast, the distribution of the entanglement has much lower variance in the analog case, although with a significantly higher mean than the noiseless value. 

In order to study a real use case, we perform an approximate matrix product state simulation of this system for $N=40$ qubits. To understand the effect of the necessary truncation, we choose a small bond dimension of $\chi = 32$, and stop the simulation whenever the estimated fidelity drops below $0.95$.
In the right panel of Fig. \ref{fig:ent} we represent the entanglement entropy curves for different trajectories as well as the noiseless trajectory. While this low bond dimension is sufficient to simulate accurately the noiseless circuit, it allows us to simulate the noisy trajectories accurately up to shorter times on average. As expected, some digital trajectories hit the fidelity cutoff very early while others can be simulated accurately for longer times. However, the overall accurate simulation time of the noisy quantum circuit is given by the shortest time among all the trajectories sampled according to \eqref{eq:sampling}. The analog trajectories displaying similar growth of entanglement hit the fidelity threshold at approximately the same time. Overall, we thus see that the analog sampling does not change significantly the maximal noisy circuit depth that can be simulated with tensor network techniques. However, the analog sampling still significantly decreases the number of trajectories to sample.

\section{Summary and discussion}
We have introduced a new technique for efficient simulation of noisy quantum computers. It consists in averaging over trajectories that are generated by systematically inserting a small perturbation after every gate, instead of rarely inserting a large perturbation, as usually performed. It offers a speedup in runtime of order $10-100$ compared to the widely used Monte-Carlo sampling over quantum trajectories, in identical settings. As such, it does not push the boundaries of what is classically simulable, but instead it helps the quantum practitioner and speeds up algorithm development.

There are a few future directions that are worth exploring. For example, we have considered in this work the consequences of this analog trajectory sampling on classical simulation. However, there are noise mitigation routines that consist in implementing on the hardware ``artificial" noise channels \cite{temme2017error,giurgica2020digital,kim2023evidence}. The analog trajectories could help reducing compilation overhead in this setting.

\subsection*{Acknowledgements}
We thank Yi Hsiang Chen, Karl Mayer and Sheng-Hsuan Lin for comments on the draft. E.G. acknowledges support by the Bavarian Ministry of Economic Affairs, Regional Development and Energy (StMWi) under project Bench-QC (DIK0425/01). K.H. and H.D. acknowledge support by the German Federal Ministry of Education and Research (BMBF) through the project
EQUAHUMO (grant number 13N16069) within the funding program quantum technologies - from basic research to market. 


%

\onecolumngrid

\appendix
\section{Proof of formula \eqref{qQ} \label{appendixa}}
Given a set of $p_S$, we are looking for a set of $q_S$ such that
\begin{equation}
   \mathcal{N}(\rho)\equiv \sum_{S'\in \mathcal{S}_M} p_{S'} S' \rho S'=\left[\prod_{S'\in\mathcal{S}_M}\mathcal{N}_{q_{S'},S'}\right](\rho)
\end{equation}
holds for all $\rho$. Let us thus set $\rho=S$ for $S\in\mathcal{S}_M$. We have $S'SS'=S$ if $S'$ commutes with $S$, and $S'SS'=-S$ if $S'$ anticommutes with $S$. Hence we have on the left-hand side
\begin{equation}
\begin{aligned}
    \mathcal{N}(S)&=\left(\sum_{S'\in c(S)}p_{S'}-\sum_{S'\in a(S)}p_{S'}\right)S\,,\\
    &=\left(1-2\sum_{S'\in a(S)}p_{S'}\right)S\,.
\end{aligned}
\end{equation}
On the right-hand side, we have
\begin{equation}
    \mathcal{N}_{q_{S'},S'}(S)=\begin{cases}
        S\,,\qquad \text{if }S'\in c(S)\\
        (1-2q_{S'})S\,,\qquad \text{if }S'\in a(S)\,.
    \end{cases}
\end{equation}
Hence we obtain for all $S\in\mathcal{S}_M$
\begin{equation}
\begin{aligned}\label{eqtor}
     1-2\sum_{S'\in a(S)}p_{S'}&=\prod_{S'\in a(S)}(1-2q_{S'})\,.
\end{aligned}
\end{equation}
    Let us introduce
\begin{equation}\label{defx}
    x=\frac{\prod_{S'\in a(S)} \prod_{S''\in a(S')}(1-2q_{S''})}{\prod_{S'\in c(S)} \prod_{S''\in a(S')}(1-2q_{S''})}\,.
\end{equation}
From \eqref{eqtor}, we have
\begin{equation}\label{xasp}
    x=\frac{\prod_{S'\in a(S)} \left(1-2\sum_{S''\in a(S')}p_{S''}\right)}{\prod_{S'\in c(S)} \left(1-2\sum_{S''\in a(S')}p_{S''}\right)}\,.
\end{equation}
We now fix $T\in\mathcal{S}_M$, and count how many times $1-2q_T$ appears in the numerator and denominator of \eqref{defx}. In the numerator (resp. denominator), $1-2q_T$ appears exactly the same number of times as the number of elements $S'\in\mathcal{S}_M$ that anticommute both with $S$ and $T$ (resp. that commute with $S$ and anticommute with $T$). Indeed, $T=S''$ if and only if $T$ anticommutes with $S'$, and by definition $S'$ anticommutes (resp. commutes) with $S$ for the numerator (resp. denominator).

Let us first treat the case $T=S$. Then, $1-2q_T$ cannot appear in the denominator (otherwise, $S$ would both commute and anticommute with the corresponding $S'$). In the numerator, $1-2q_T$ appears the same number of times as the number of strings $S'\in\mathcal{S}_M$ that anticommute with $S$. Since $S\neq I$, there exists at least one string $S^{\prime *}\in a(S)$ that anticommutes with $S$. Then, since strings either commute or anticommute among themselves, we have $S'\in c(S)$ if and only if $S'S^{\prime *}\in a(S)$. As a consequence, $c(S)$ and $a(S)$ must have the same number of elements, and so since their sum is equal to $4^M$, each of them has $4^M/2$ elements. Hence, $1-2q_T$ appears $4^M/2$ times in the numerator of \eqref{defx}, and $0$ times in the denominator.

Let us now treat the case $T\neq S$. A string $S'$ anticommutes with $T$ and commutes with $S$ if and only if $S'$ anticommutes with both $T$ and $ST$. Hence, the number of strings $S'$ that anticommute with $T$ and commute with $S$ is equal to the number of strings $S'$ that anticommute both with $T$ and $ST$. Let us now take two strings $S_1\neq S_2$ different from identity. There has to be at least one element $S_1^*$ (resp. $S_2^*$) that anticommute with $S_1$ (resp. $S_2$), and since $S_1\neq S_2$ one can impose $S_1^*\neq S_2^*$. Then, a string $S'$ has some commutation relations with $S_1,S_2$ if and only if the string $S_{1,2}^*S'$ has the opposite commutation relation with $S_{1,2}$ but the same with $S_{2,1}$, if and only if the string $S_{1}^*S_{2}^*S'$ has the opposite commutation relation with both $S_1$ and $S_2$. Hence there has to be the same number of strings with these $4$ possible commutation relations with $S_1,S_2$. Hence, there are always exactly $4^M/4$ strings $S'$ that anticommute with both $S_1$ and $S_2$ if $S_1\neq S_2$ are different from identity. There are also always exactly $4^M/4$ strings $S'$ that anticommute with $S_1$ and commute with $S_2$ if $S_1\neq S_2$ are different from identity. As a consequence, $1-2q_T$ appears the same number of times ($4^M/4$ times) in the numerator and denominator of \eqref{defx}. 

It follows that we have
\begin{equation}
    x=(1-2q_S)^{4^M/2}\,.
\end{equation}
Inverting this relation and using \eqref{xasp}, we obtain the result \eqref{qQ}. 

\section{Derivation of coherent noise channel\label{app:coh}}
We consider the noise channel
\begin{equation}
    \rho\mapsto (1-q)\rho+q e^{i\alpha X}\rho e^{-i\alpha X}\,.
\end{equation}
For this to be equivalent to 
\begin{equation}
    \rho\mapsto \mathbb{E}[e^{i\theta X}\rho e^{-i\theta X}]\,,
\end{equation}
where the angle $\theta$ is drawn with probability density $f(\theta)$, one requires
\begin{equation}
    \begin{aligned}
       & \int \cos^2(\theta)f(\theta)\D{\theta}=(1-q)+q\cos^2\alpha\\
       &\int \sin^2(\theta)f(\theta)\D{\theta}=q \sin^2\alpha\\
        &\int \cos(\theta)\sin(\theta)f(\theta)\D{\theta}=q \cos\alpha \sin\alpha\,.
    \end{aligned}
\end{equation}
We take the following Gaussian ansatz
\begin{equation}
    f(\theta)=\frac{e^{-\frac{(\theta-\mu)^2}{2\sigma^2}}}{\sqrt{2\pi \sigma^2}}\,,
\end{equation}
for some $\sigma,\mu$ to be expressed in terms of $q,\alpha$. We have
\begin{equation}
    \begin{aligned}
        \int \sin^2(\theta)f(\theta)\D{\theta}&=\frac{1}{2}-\frac{e^{-2\sigma^2}}{2}\cos(2\mu)\\
        \int \cos(\theta)\sin(\theta)f(\theta)\D{\theta}&=\frac{1}{2}e^{-2\sigma^2}\sin(2\mu)\,.
    \end{aligned}
\end{equation}
These equations are then readily inverted to find $\sigma(q,\alpha)$ and $\mu(q,\alpha)$ given in the main text.

\section{Angle distributions \label{sec:distributionangle}}
We provide here the different angle distributions studied in Figure \ref{fig:analog}. They are parameterized by a scalar $a>0$ that must be chosen so that $\int_{-\infty}^\infty \sin^2\theta f(\theta)\D{\theta}=q$ with $q$ a given probability. When an explicit analytic expression is available, we provide it alongside the angle distribution. Otherwise, the corresponding value of $a$ has to be computed numerically.
\begin{itemize}
    \item Gaussian
    \begin{equation}
        f(\theta)=\frac{1}{\sqrt{2\pi a^2}}e^{-\frac{\theta^2}{2a^2}}\,. \qquad\qquad a=\sqrt{-\frac{1}{2}\log(1-2q)}\,.
    \end{equation}
    \item discrete
    \begin{equation}
        f(\theta)=\frac{1}{2}(\delta(\theta-a)+\delta(\theta+a))\,. \qquad\qquad a=\arcsin\sqrt{q}\,.
    \end{equation}
    \item uniform
    \begin{equation}
        f(\theta)=\frac{1}{2a}\,,\qquad -a\leq\theta\leq a\,.
    \end{equation}
    \item exponential
    \begin{equation}
        f(\theta)=\frac{1}{2a}e^{-\frac{|\theta|}{a}}\,. \qquad\qquad a=\sqrt{\frac{q}{4-2q}}\,.
    \end{equation}
    \item Cauchy
    \begin{equation}
        f(\theta)=\frac{1}{\pi}\frac{a}{x^2+a^2}\,. \qquad\qquad a=-\frac{1}{2}\log(1-2q)\,.
    \end{equation}
    \item semi-circular
    \begin{equation}
        f(\theta)=\frac{2}{\pi a^2}\sqrt{a^2-\theta^2}\,,\qquad -a\leq\theta\leq a\,.
    \end{equation}
    \item raised cosine
    \begin{equation}
        f(\theta)=\frac{1}{2a}\cos(\frac{\theta \pi}{a})\,,\qquad -a\leq\theta\leq a\,.
    \end{equation}
\end{itemize}

\section{Tutorial---Example application and comparison of analog and digital sampling\label{sec:samplecode}}

\definecolor{codegreen}{rgb}{0,0.6,0}
\definecolor{codegray}{rgb}{0.5,0.5,0.5}
\definecolor{codepurple}{rgb}{0.58,0,0.82}
\definecolor{backcolour}{rgb}{0.95,0.95,0.92}

\lstdefinestyle{mystyle}{
    backgroundcolor=\color{backcolour},   
    commentstyle=\color{codegreen},
    keywordstyle=\color{magenta},
    numberstyle=\tiny\color{codegray},
    stringstyle=\color{codepurple},
    basicstyle=\ttfamily\footnotesize,
    breakatwhitespace=false,         
    breaklines=true,                 
    captionpos=b,                    
    keepspaces=true,                 
    numbers=left,                    
    numbersep=5pt,                  
    showspaces=false,                
    showstringspaces=false,
    showtabs=false,                  
    tabsize=2
}
\lstset{style=mystyle}

\begin{figure}
    \centering
    \includegraphics[width=\linewidth]{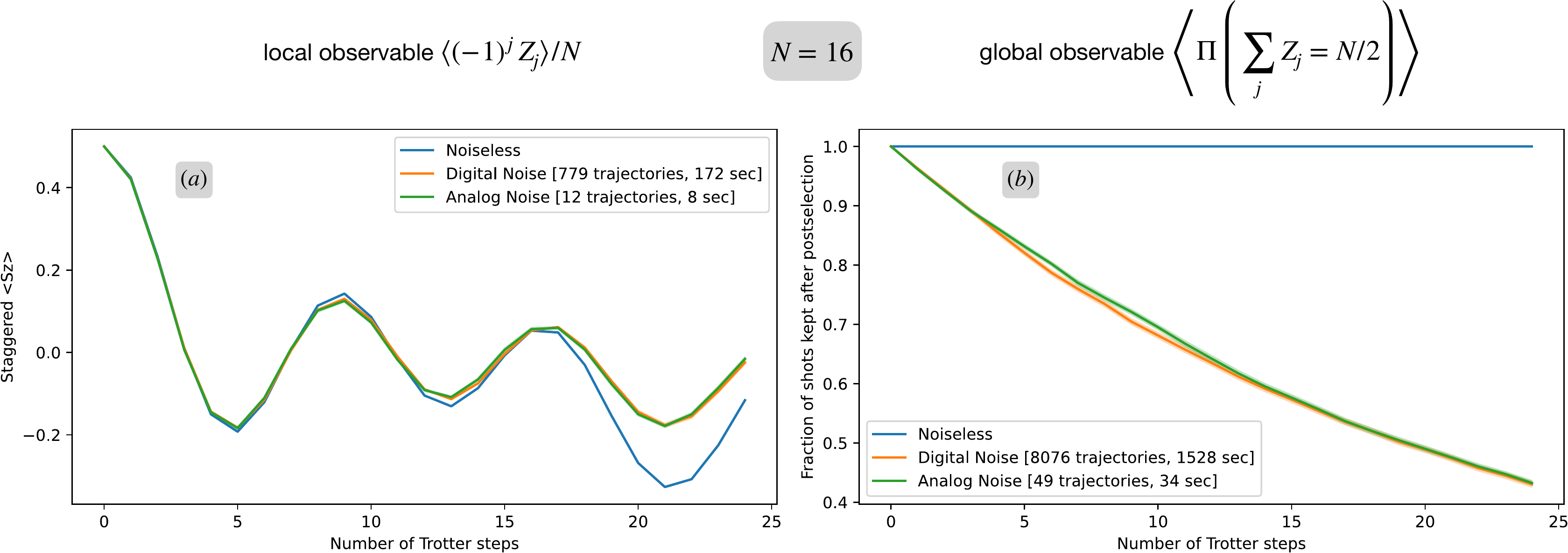}
    \caption{Sanity check and relative performance of digital vs. analog simulation of noisy circuits. a) A $N=16$ 1D XY model is initialised in a product state and the $\pi$-staggered $Z$-magnetisation is computed during a quench. The runtime speedup for the analog simulations is roughly a factor 20. b) Assessing the feasibility of postselecting based on the conserved quantity $\sum_j Z_j$, by computing noisy expectation values of the projector into the subspace that is retained under filtering. Runtime speedup for analog simulations is roughly a factor 45. Trajectories are added until the maximum standard error on the mean is less than 0.005.}
    \label{fig_example}
\end{figure}

In this section we will walk through an example application of noisy quantum circuit simulation and use example code listings that can be copied and pasted into the reader's own applications. The example we have in mind is the simulation of the quench dynamics of the 1D XY model with $N=16$ spins and periodic boundary conditions:
\begin{equation}
    H = \sum_{i=1}^N X_i X_{i+1} + Y_i Y_{i+1}.
\end{equation}
First, we are interested in the evolution of the staggered $Z$-magnetisation $S^z(\pi) =  \sum_i (-1)^i Z_i/N$ after initialising the state in a computational basis state with a down-spin at every fourth site, i.e. $\ket{\psi_0} = \ket{00010001 \dots}$ and $\braket{\psi_0 | S^z(\pi)| \psi_0} = 1/2$. The evolution is done using $m = 0, 1, \dots 25$ first-order Trotter steps, i.e. $\ket{\psi(m)} = U^m \ket{\psi_0}$ and $U = \prod_{i=1}^N \exp(-i \tau Y_i Y_{i+1}) \exp(-i \tau X_i X_{i+1})$. We choose $\tau = 0.1$.

We will first establish the ground truth by  obtaining the noiseless values as follows:
\begin{lstlisting}[language=Python,numbers=none]
from qiskit import QuantumCircuit
from qiskit.quantum_info import SparsePauliOp
from qiskit_aer import AerSimulator
import numpy as np
from matplotlib import pyplot as plt

##################################
####  Some helper functions   ####
##################################
def get_statevector(qc, simulator, noise_model=None):
    qc_temp = qc.copy()
    qc_temp.save_statevector()
    sv = simulator.run(qc_temp, noise_model=noise_model, shots=1).result().data()['statevector']
    return sv
def trotter_step(tau, couplings):
    N = max(list(map(max, couplings))) + 1
    qc = QuantumCircuit(N)
    for coupling in couplings:
        qc.rxx(tau * 2, coupling[0], coupling[1])
        qc.ryy(tau * 2, coupling[0], coupling[1])
    return qc

##################################
####     Set up parameters    ####
##################################
N=16
tau = 0.1
N_timesteps = 25
couplings = [(i,(i+1)%N) for i in range(N)]
magnetization=sum([SparsePauliOp('I'*x + 'Z' + 'I'*(N-x-1),coeffs=[(-1)**x]) for x in range(N)])/N
simulator = AerSimulator()

##################################
#### Obtain noiseless values  ####
##################################
qc = QuantumCircuit(N)
for j in range(1,N,4): qc.x(j)
statevector = get_statevector(qc, simulator)
expectation_value = np.real(statevector.reverse_qargs().expectation_value(magnetization))
expectation_values = [expectation_value]
for t in range(1,N_timesteps):
    qc = QuantumCircuit(N)
    qc.set_statevector(statevector)
    qc = qc.compose(trotter_step(tau, couplings))
    statevector = get_statevector(qc, simulator)
    expectation_value = np.real(statevector.reverse_qargs().expectation_value(magnetization))
    expectation_values.append(expectation_value)

ts = np.arange(N_timesteps)
fig, (ax1, ax2) = plt.subplots(1,2, figsize=(14,4))
ax1.plot(ts, expectation_values,'C0', label='Noiseless')
\end{lstlisting}
Let us make two side remarks before moving on to the noisy simulations: First, we are using exact expectation value computations rather than sampling from the statevector to estimate observables. While many online tutorials use sampling-based estimators to prepare users for running on real devices, it should go without saying that shot-based methods should almost never be used for experiment design on classical computers. Second, naive simulations of the timeseries $m=0,1, \dots M_\mathrm{max}$ will take $M_\mathrm{max} (M_\mathrm{max}+1)/2 = \mathcal{O}(M_\mathrm{max}^2)$ time which can be reduced to linear time $\mathcal{O}(M_\mathrm{max})$ by keeping track of the statevector during the evolution as shown above.

We would now like to simulate the performance of a noisy quantum computer for the same task. To this end we equip each of our two-qubit $R_{xx}$ and $R_{yy}$ gates with depolarising noise $2\times 10^{-3}$. We will keep sampling trajectories until we obtain a maximum standard error on the mean of $\sigma = 0.005$. We first solve this task using standard noise model implementations in qiskit:
\begin{lstlisting}[language=Python,numbers=none]
###########################################
#### Digital noisy simulations         ####
#### for the staggered magnetisation   ####
###########################################
from qiskit_aer.noise import NoiseModel, depolarizing_error
from time import time

simulator = AerSimulator()
noise_model = NoiseModel()
epsilon = 2e-3
error_2 = depolarizing_error(epsilon, 2)
noise_model.add_all_qubit_quantum_error(error_2, ['rxx', 'ryy'])
print(noise_model.to_dict())
desired_accuracy = 0.005
t0 = time()
error = 0
n = 0
while n<5 or error > desired_accuracy:
    n=n+1
    qc = QuantumCircuit(N)
    for j in range(1, N, 4): qc.x(j)
    statevector = get_statevector(qc, simulator, noise_model=noise_model)
    trajectory_values = [np.real(statevector.reverse_qargs().expectation_value(magnetization))]
    for t in range(1, N_timesteps):
        qc = QuantumCircuit(N)
        qc.set_statevector(statevector)
        qc = qc.compose(trotter_step(tau, couplings))
        statevector = get_statevector(qc, simulator, noise_model=noise_model)
        expectation_value = np.real(statevector.reverse_qargs().expectation_value(magnetization))
        trajectory_values.append(expectation_value)
    if n==1: expectation_values_digital = np.array(trajectory_values).reshape((1,N_timesteps))
    else: expectation_values_digital = np.r_[expectation_values_digital, [np.array(trajectory_values)]]
    std_digital   = np.std(expectation_values_digital,axis=0)/np.sqrt(n)
    error = max(std_digital)
    print('Digital Trajectory #{}. Standard error on the mean = {:.3f}'.format(n, error))

means_digital = np.mean(expectation_values_digital,axis=0)
t1 = time()
t_digital = t1-t0

ax1.plot(ts, means_digital, color='C1', label='Digital Noise [{} trajectories, {:.0f} sec]'.format(n, t_digital))
ax1.fill_between(ts, means_digital-std_digital, means_digital+std_digital, color='C1', alpha=0.2)
\end{lstlisting}
Second, we we use the analog strategy that we proposed in the main text of this work.
\begin{lstlisting}[language=Python,numbers=none]
###########################################
#### Analog noisy simulations          ####
#### for the staggered magnetisation   ####
###########################################
def trotter_step_w_analog_noise(tau, couplings, epsilon):
    N = max(list(map(max, couplings))) + 1
    sigma = np.sqrt(-1/16 * np.log(1-epsilon))
    qc = QuantumCircuit(N)
    for coupling in couplings:
        qc.rxx(tau * 2, coupling[0], coupling[1])
        add_noise(qc, coupling, sigma)
        qc.ryy(tau * 2, coupling[0], coupling[1])
        add_noise(qc, coupling, sigma)
    return qc
def add_noise(qc, coupling, sigma):
    #############################
    # Adding analog rotations   #
    #############################
    qc.rx(-2 * np.random.normal(scale=sigma), coupling[0])
    qc.ry(-2 * np.random.normal(scale=sigma), coupling[0])
    qc.rz(-2 * np.random.normal(scale=sigma), coupling[0])
    qc.rx(-2 * np.random.normal(scale=sigma), coupling[1])
    qc.rxx(-2 * np.random.normal(scale=sigma), coupling[0], coupling[1])
    qc.s(coupling[0])
    qc.rxx(-2 * np.random.normal(scale=sigma), coupling[0], coupling[1])
    qc.sdg(coupling[0])
    qc.h(coupling[0])
    qc.rxx(-2 * np.random.normal(scale=sigma), coupling[0], coupling[1])
    qc.h(coupling[0])
    qc.ry(-2 * np.random.normal(scale=sigma), coupling[1])
    qc.s(coupling[1])
    qc.rxx(-2 * np.random.normal(scale=sigma), coupling[0], coupling[1])
    qc.sdg(coupling[1])
    qc.ryy(-2 * np.random.normal(scale=sigma), coupling[0], coupling[1])
    qc.h(coupling[0])
    qc.s(coupling[0])
    qc.ryy(-2 * np.random.normal(scale=sigma), coupling[0], coupling[1])
    qc.sdg(coupling[0])
    qc.h(coupling[0])
    qc.rz(-2 * np.random.normal(scale=sigma), coupling[1])
    qc.h(coupling[0])
    qc.rzz(-2 * np.random.normal(scale=sigma), coupling[0], coupling[1])
    qc.h(coupling[0])
    qc.s(coupling[0])
    qc.h(coupling[0])
    qc.rzz(-2 * np.random.normal(scale=sigma), coupling[0], coupling[1])
    qc.h(coupling[0])
    qc.sdg(coupling[0])
    qc.rzz(-2 * np.random.normal(scale=sigma), coupling[0], coupling[1])
    return qc



t0 = time()
n = 0
while n<5 or error > desired_accuracy:
    n=n+1
    qc = QuantumCircuit(N)
    for j in range(1, N, 4): qc.x(j)
    statevector = get_statevector(qc, simulator)
    trajectory_values = [np.real(statevector.reverse_qargs().expectation_value(magnetization))]
    for t in range(1, N_timesteps):
        qc = QuantumCircuit(N)
        qc.set_statevector(statevector)
        qc = qc.compose(trotter_step_w_analog_noise(tau, couplings, epsilon))
        statevector = get_statevector(qc, simulator)
        expectation_value = np.real(statevector.reverse_qargs().expectation_value(magnetization))
        trajectory_values.append(expectation_value)
    if n==1: expectation_values_analog = np.array(trajectory_values).reshape((1,N_timesteps))
    else: expectation_values_analog = np.r_[expectation_values_analog, [np.array(trajectory_values)]]
    std_analog   = np.std(expectation_values_analog,axis=0)/np.sqrt(n)
    error = max(std_analog)
    print('Analog Trajectory #{}. Standard error on the mean = {:.3f}'.format(n, error))

means_analog = np.mean(expectation_values_analog,axis=0)
t1 = time()
t_analog = t1-t0
ax1.plot(ts, means_analog, color='C2', label='Analog Noise [{} trajectories, {:.0f} sec]'.format(n, t_analog))
ax1.fill_between(ts, np.array(means_analog)-np.array(std_analog), np.array(means_analog)+np.array(std_analog), color='C2', alpha=0.2)
ax1.legend(loc='best')
\end{lstlisting}
The comparison of the digital and analog strategies for the staggered magnetisation is shown in Fig.~\ref{fig_example}a. Despite the analog strategy taking longer per trajectory (since more two-qubit gates are applied), it requires much less trajectories to reach the prescribed precision and so comes out roughly a factor 20 ahead in overall runtime.

\begin{figure}
    \centering
    \includegraphics[width=\linewidth]{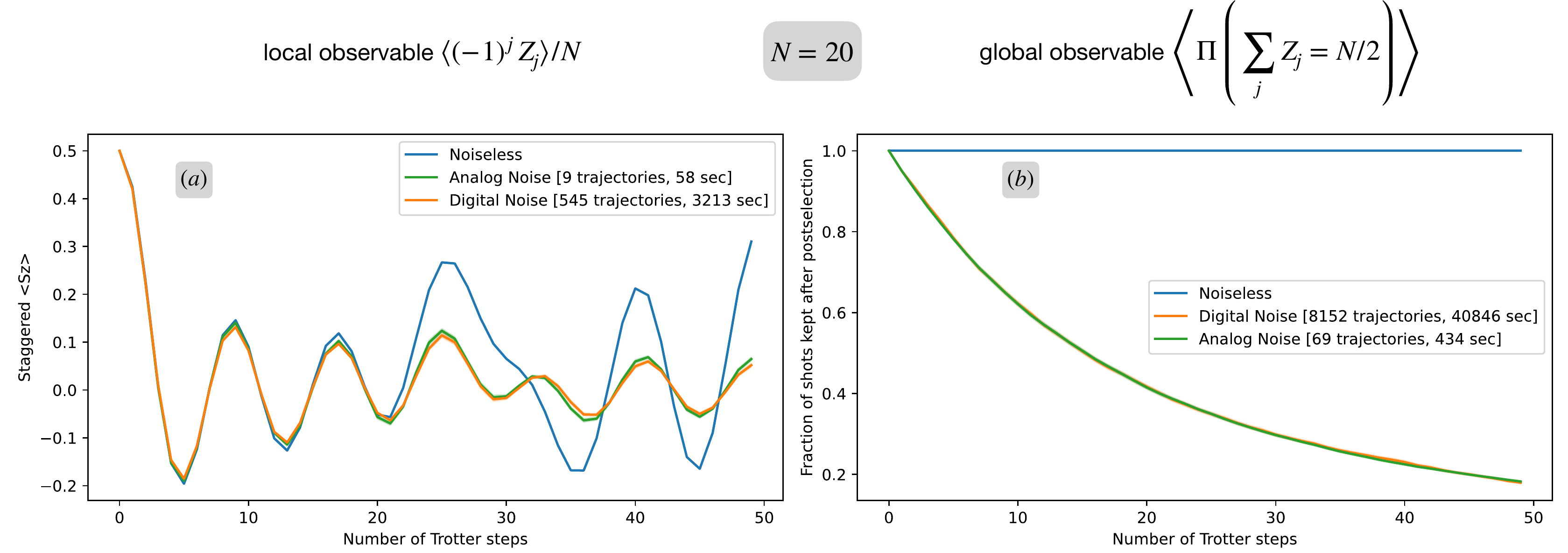}
    \caption{Same as Figure \ref{fig_example} but for a larger system size $N=20$ and more (50) Trotter steps.}
    \label{fig_example2}
\end{figure}

In the second part of our application we are switching focus from a local to a global observable: One commonly used technique to error mitigate outcomes of a noisy quantum computer is to use symmetry filtering. In this framework, one identifies symmetries of the initial state and noiseless evolution and postselects shots that fulfil corresponding symmetry constraints. For example, our noiseless evolution in this section fulfils $\sum_j Z_j = N/2$. We now want to answer the question ``how many shots do we expect to retain on average after such a symmetry postselection in the present scenario?". Similar to the magnetisation estimates before, this question could be answered using a shot-based apprach, albeit in a very inefficient manner. It is much better to compute the expectation value of the projector $\Pi$ onto the correct symmetry sector, i.e.
\begin{align}
    \Pi = \begin{cases}
			1, & \text{if $\sum_j Z_j=N/2$}\\
            0, & \text{otherwise}.
		 \end{cases}
\end{align}
The noiseless computation proceeds as follows:
\begin{lstlisting}[language=Python,numbers=none]
#####################################
#### Noiseless values of           ##
#### projector on Sz = 1/2 sector  ##
#### (should be = 1 for all times) ##
#####################################
szvec = np.array([N - 2 * i.bit_count() for i in np.arange(2 ** N)], dtype=np.byte)
truth = (szvec == int(N * (1 - 2 * 1/4)))
qc = QuantumCircuit(N)
for j in range(1,N,4): qc.x(j)
statevector = get_statevector(qc, simulator)
expectation_value = sum(np.abs(np.array(statevector)[truth]) ** 2)
expectation_values = [expectation_value]
for t in range(1,N_timesteps):
    qc = QuantumCircuit(N)
    qc.set_statevector(statevector)
    qc = qc.compose(trotter_step(tau, couplings))
    statevector = get_statevector(qc, simulator)
    expectation_value = sum(np.abs(np.array(statevector)[truth]) ** 2)
    expectation_values.append(expectation_value)
ax2.plot(ts, expectation_values,'C0', label='Noiseless')
\end{lstlisting}
The digital approach is given in the following snippet
\begin{lstlisting}[language=Python,numbers=none]
#######################################
#### Digital Noisy Simulations      ###
#### of the projector onto Sz=N/2   ###
#######################################
t0 = time()
error = 0
n = 0
while n<5 or error > desired_accuracy:
    n=n+1
    qc = QuantumCircuit(N)
    for j in range(1, N, 4): qc.x(j)
    statevector = get_statevector(qc, simulator, noise_model=noise_model)
    trajectory_values = [sum(np.abs(np.array(statevector)[truth])**2)]
    for t in range(1, N_timesteps):
        qc = QuantumCircuit(N)
        qc.set_statevector(statevector)
        qc = qc.compose(trotter_step(tau, couplings))
        statevector = get_statevector(qc, simulator, noise_model=noise_model)
        expectation_value = sum(np.abs(np.array(statevector)[truth])**2)
        trajectory_values.append(expectation_value)
    if n==1: expectation_values_digital = np.array(trajectory_values).reshape((1,N_timesteps))
    else: expectation_values_digital = np.r_[expectation_values_digital, [np.array(trajectory_values)]]
    std_digital   = np.std(expectation_values_digital,axis=0)/np.sqrt(n)
    error = max(std_digital)
    print('Digital Trajectory #{}. Standard error on the mean = {:.3f}'.format(n, error))

means_digital = np.mean(expectation_values_digital,axis=0)
t1 = time()
t_digital = t1-t0

ax2.plot(ts, means_digital, color='C1', label='Digital Noise [{} trajectories, {:.0f} sec]'.format(n, t_digital))
ax2.fill_between(ts, means_digital-std_digital, means_digital+std_digital, color='C1', alpha=0.2)
\end{lstlisting}
Finally, the analog approach is as follows:
\begin{lstlisting}[language=Python,numbers=none]
#####################################
#### Analog way of computing       ##
#### projector on Sz = 1/2 sector  ##
#### & final plotting              ##
#####################################
t0 = time()
error = 0
n = 0
while n<5 or error > desired_accuracy:
    n=n+1
    qc = QuantumCircuit(N)
    for j in range(1, N, 4): qc.x(j)
    statevector = get_statevector(qc, simulator)
    trajectory_values = [sum(np.abs(np.array(statevector)[truth])**2)]
    for t in range(1, N_timesteps):
        qc = QuantumCircuit(N)
        qc.set_statevector(statevector)
        qc = qc.compose(trotter_step_w_analog_noise(tau, couplings, epsilon))
        statevector = get_statevector(qc, simulator)
        expectation_value = sum(np.abs(np.array(statevector)[truth])**2)
        trajectory_values.append(expectation_value)
    if n==1: expectation_values_analog = np.array(trajectory_values).reshape((1,N_timesteps))
    else: expectation_values_analog = np.r_[expectation_values_analog, [np.array(trajectory_values)]]
    std_analog   = np.std(expectation_values_analog,axis=0)/np.sqrt(n)
    error = max(std_analog)
    print('Analog Trajectory  #{}. Standard error on the mean = {:.3f}'.format(n, error))

means_analog = np.mean(expectation_values_analog,axis=0)
t1 = time()
t_analog = t1-t0

ax2.plot(ts, means_analog, color='C2', label='Analog Noise [{} trajectories, {:.0f} sec]'.format(n, t_analog))
ax2.fill_between(ts, means_analog-std_analog, means_analog+std_analog, color='C2', alpha=0.2)
ax2.legend(loc='best')

ax1.set_xlabel('Number of Trotter steps')
ax1.set_ylabel('<Staggered Sz>')
ax2.set_xlabel('Number of Trotter steps')
ax2.set_ylabel('Fraction of shots kept after postselection')

plt.tight_layout()
plt.savefig('plot.pdf')
\end{lstlisting}
The results of this study are shown in Fig.~\ref{fig_example}b. In the case of a global observable, the advantage of the analog technique is even more extreme than in the local case showing a x45 runtime speedup versus the standard digital approach. We have encountered similar behaviour for other local observables, like fidelities with given target states or overlaps with computational basis states, with even larger runtime gains for larger system sizes and longer evolutions. For example, scaling up the present simulation to $N=20$ and 50 Trotter steps, increases the speedup to a factor 55 for the local observable and a factor 94 for the global observable, as shown in Figure ~\ref{fig_example2}.
Runtime measurements were taken on an M2 MacBook Air with 8GB memory.

\end{document}